\documentclass[pra,twocolumn,groupedaddress,superscriptaddress,amsmath,amssymb,a4paper]{revtex4-2}
\usepackage{geometry}
\usepackage{graphicx}
\usepackage{mathrsfs}
\usepackage{amssymb}
\usepackage{amsmath}
\usepackage{amsthm}
\usepackage{amsbsy}
\usepackage{bm}
\usepackage{hyperref}

\theoremstyle{definition}

\newcommand{\bra}[1]{\langle #1|}
\newcommand{\ket}[1]{| #1 \rangle }

\begin{document}

\title{Multipartite entanglement to boost superadditivity of coherent information\\ in quantum communication lines with polarization dependent losses}

\author{Sergey N. Filippov}

\affiliation{Steklov Mathematical Institute of Russian Academy of
Sciences, Gubkina St. 8, Moscow 119991, Russia}

\begin{abstract}
Coherent information quantifies the achievable rate of the
reliable quantum information transmission through a communication
channel. Use of the correlated quantum states instead of the
factorized ones may result in an increase in the coherent
information, a phenomenon known as superadditivity. However, even
for simple physical models of channels it is rather difficult to
detect the superadditivity and find the advantageous multipartite
states. Here we consider the case of polarization dependent losses
and propose some physically motivated multipartite entangled
states which outperform all factorized states in a wide range of
the channel parameters. We show that in the asymptotic limit of
the infinite number of channel uses the superadditivity phenomenon
takes place whenever the channel is neither degradable nor
antidegradable. Besides the superadditivity identification, we
also provide a method how to modify the proposed states and get a
higher quantum communication rate by doubling the number of
channel uses. The obtained results give a deeper understanding of
coherent information in the multishot scenario and may serve as a
benchmark for quantum capacity estimations and future approaches
toward an optimal strategy to transfer quantum information.
\end{abstract}

\maketitle


\section{Introduction} \label{section-introduction}

Quantum information represents quantum states in a variety of
forms including superpositions and entanglement. Quantum
information significantly differs from classical information
because quantum states cannot be deterministically cloned in
contrast to classical letters. On the other hand, it is quantum
information that should be transferred along physical
communication lines to connect quantum computers in a network and
manipulate a long-distance entanglement, which potentially has
numerous applications~\cite{horodecki-2009,erhard-2020}. A
successful transmission of quantum information through a noisy
channel implies a perfect transfer (in terms of the fidelity) of
any quantum state by arranging appropriate encoding and decoding
procedures at the input and the output of the channel,
respectively, see
Refs.~\cite{lloyd-1997,barnum-1998,shor-2002,devetak-2005}.
Physical meaning of quantum information transfer is also discussed
in Ref.~\cite{wilde-2013} from the viewpoint of creating
entanglement between the apart laboratories, provided the channel
can be used many times. A multishot scenario implies $n$ uses of
the communication channel so $n$ quantum information carriers,
e.g., photons, are treated as a whole. By $\varrho^{(n)}$ denote
the average density operator of an ensemble of $n$-partite states
used in the quantum communication task~\cite{lloyd-1997}. In this
paper, we report entangled $n$-partite states $\varrho^{(n)}$ that
enable to transmit an increasing amount of quantum information
with the increase of $n$.

If each of $n$ information carriers propagates through a
memoryless noisy quantum channel $\Phi$, then the average noisy
output is $\Phi^{\otimes n}[\varrho^{(n)}]$. The decoder aims at
reproducing the encoded state. A figure of merit for this task is
the achievable communication rate that quantifies how many qubits
per channel use can be reliably transmitted in the sense that the
error vanishes in the asymptotic limit of infinitely many channel
uses. The quantum capacity $Q(\Phi)$ is defined as the supremum of
achievable communication rates among all possible encodings and
decodings. The result of the seminal paper~\cite{devetak-2005}
generalizes some previous
observations~\cite{lloyd-1997,barnum-1998,shor-2002} and shows
that
\begin{equation*}
Q(\Phi) = \lim_{n \rightarrow \infty} Q_n(\Phi),
\end{equation*}

\noindent where
\begin{eqnarray*}
&& Q_n(\Phi) = \frac{1}{n} Q_1(\Phi^{\otimes n}), \\
&& Q_1(\Psi) = \sup_{\varrho} I_c(\varrho,\Psi), \\
&& I_c(\varrho,\Psi) = S(\Psi[\varrho]) -
S(\widetilde{\Psi}[\varrho]).
\end{eqnarray*}

\noindent $I_c(\varrho,\Psi)$ is a so-called coherent information
that quantifies an asymmetry between the von Neumann entropy
$S(\Psi[\varrho])$ of the channel output and the von Neumann
entropy $S(\widetilde{\Psi}[\varrho])$ of a complementary channel
output. In other words, the coherent information effectively
quantifies an asymmetry between the receiver information
$S(\Psi[\varrho])$ and the information
$S(\widetilde{\Psi}[\varrho])$ diluted into the environment. To
make this description precise, consider a quantum channel
$\Psi:{\cal B}({\cal H}_A) \rightarrow {\cal B}({\cal H}_B)$,
where ${\cal H}_A$ and ${\cal H}_B$ are the Hilbert spaces of
input and output, respectively, and ${\cal B}({\cal H})$ denotes a
set of bounded operators on ${\cal H}$. Hereafter, we consider
finite-dimensional Hilbert spaces because we will further focus on
a finite-dimensional physical model of polarization dependent
losses. The Stinespring dilation for $\Psi:{\cal B}({\cal H}_A)
\rightarrow {\cal B}({\cal H}_B)$ reads as follows in the
Schr\"{o}dinger picture:
\begin{equation} \label{stinespring-Phi}
\Psi[\varrho] = {\rm tr}_E \big[ V \varrho V^{\dag} \big],
\end{equation}

\noindent where $V: {\cal H}_A \rightarrow {\cal H}_B \otimes
{\cal H}_E$ is an isometry ($V^{\dag}V = I_A$), ${\cal H}_E$
denotes the Hilbert space of the effective environment, and ${\rm
tr}_E$ is the partial trace with respect to the effective
environment (see, e.g.,~\cite{holevo-2012}). The formula
\begin{equation*}
\widetilde{\Psi}[\varrho] = {\rm tr}_B \big[ V \varrho V^{\dag}
\big]
\end{equation*}

\noindent defines a channel $\widetilde{\Psi}:{\cal B}({\cal H}_A)
\rightarrow {\cal B}({\cal H}_E)$ that is complementary to
$\Psi:{\cal B}({\cal H}_A) \rightarrow {\cal B}({\cal H}_B)$.
Since the Stinespring dilation~\eqref{stinespring-Phi} is not
unique for a given channel $\Psi$, neither is the complementary
channel $\widetilde{\Psi}$; however, all complementary channels
are isometrically equivalent (see, e.g.,~\cite{holevo-2012}).

Suppose two quantum channels $\Phi:{\cal B}({\cal H}_A)
\rightarrow {\cal B}({\cal H}_B)$ and $\Phi':{\cal B}({\cal
H}_{A'}) \rightarrow {\cal B}({\cal H}_{B'})$ are both degradable,
i.e., there exist quantum channels ${\cal D}$ and ${\cal D}'$ such
that $\widetilde{\Phi} = {\cal D} \circ {\Phi}$ and
$\widetilde{\Phi}' = {\cal D}' \circ {\Phi}'$; the symbol $\circ$
denotes a concatenation of maps. Then the coherent information is
subadditive~\cite{devetak-shor-2005} in the sense that
\begin{equation} \label{coh-inf-subadditivity}
I_c(\varrho_{AA'}, \Phi \otimes \Phi') \leq I_c(\varrho_{A}, \Phi)
+ I_c(\varrho_{A'}, \Phi').
\end{equation}

\noindent An immediate consequence of
Eq.~\eqref{coh-inf-subadditivity} is the additivity of the
one-shot capacity, $Q_1(\Phi \otimes \Phi') = Q_1(\Phi) +
Q_1(\Phi')$. If $\Phi' = \Phi^{\otimes (n-1)}$, then we get
$Q_1(\Phi^{\otimes n}) = n Q_1(\Phi)$ by mathematical induction.
Hence, if the channel $\Phi$ is degradable, then the quantum
capacity $Q(\Phi)$ coincides with the one-shot quantum capacity
$Q_1(\Phi)$. Subadditivity of coherent information for degradable
channels significantly simplifies calculations of the quantum
capacity and shows that the quantum capacity can be achieved with
the use of classical-inspired random subspace codes of block
length 1~\cite{lloyd-1997,barnum-1998,shor-2002,devetak-2005}.

If the channel $\Phi$ is antidegradable, i.e., there exists a
quantum channel ${\cal A}$ such that $\Phi = {\cal A} \circ
\widetilde{\Phi}$, then $I_c(\varrho,\Phi)$ is nonpositive and
vanishes for pure states $\varrho = \ket{\psi}\bra{\psi}$.
Similarly, $I_c(\varrho,\Phi^{\otimes n}) \leq 0$. This implies
the trivial equality $Q(\Phi) = Q_1(\Phi) = 0$, i.e., all
encodings are equally useless for quantum information
transmission.

If $\Phi:{\cal B}({\cal H}_A) \rightarrow {\cal B}({\cal H}_B)$ is
neither degradable nor antidegradable, then it may happen that
there exists an $n$-partite quantum state $\varrho^{(n)} =
\varrho_{A_1 \ldots A_n}$ such that
\begin{equation*}
I_c(\varrho_{A_1 \ldots A_n}, \Phi^{\otimes n}) > \sum_k
I_c(\varrho_{A_k}, \Phi)
\end{equation*}

\noindent and $Q_n(\Phi) > Q_1(\Phi)$. This case corresponds to
superadditivity of coherent information, which implies that some
special quantum codes (for which $\varrho^{(n)}$ is correlated)
can outperform conventional ones (for which $\varrho^{(n)} =
(\varrho^{(1)})^{\otimes n}$). The superadditivity phenomenon is
predicted for qubit depolarizing channels if $n \geq
3$~\cite{divincenzo-1998,fern-2008}, so-called dephrasure qubit
channels if $n \geq 2$~\cite{leditzky-2018} (for which
superadditivity was also analyzed experimentally~\cite{yu-2020}),
a concatenation of an erasure qubit channel with an amplitude
damping qubit channel~\cite{siddhu-2021}, some qutrit channels and
their higher-dimensional
generalizations~\cite{siddhu-nov-2020,leditzky-2022}, and a
collection of specific channels if $n \geq n_0$, where $n_0 \geq
2$ can be arbitrary~\cite{cubitt-2015}. In this paper, we focus on
quantum communication lines with polarization dependent
losses~\cite{gisin-1997,kirby-2019,li-2018,filippov-QP-2021,filippov-quanta-2021},
which also exhibit the coherent information superadditivity for
some values of attenuation factors~\cite{filippov-2021}.

\begin{figure}
\includegraphics[width=8cm]{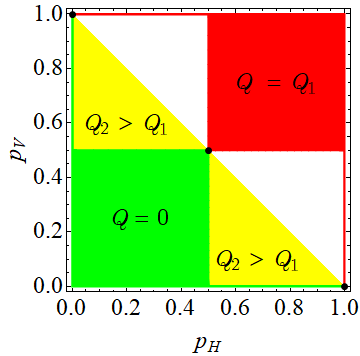}
\caption{\label{figure-1} Dimensionless attenuation factors $p_H$
and $p_V$ for horizontally and vertically polarized photons for
which the quantum channel~\eqref{pdl-channel} is degradable [red
(dark gray) area], antidegradable [green (medium gray) area], both
degradable and antidegradable (black points). Yellow (light gray)
regions correspond to the coherent-information superadditivity
detected with the use of two-letter
encodings~\cite{filippov-2021}.}
\end{figure}

Consider a lossy quantum communication line such that the
transmission coefficient for horizontally polarized photons,
$p_H$, differs from that for vertically polarized photons, $p_V$.
The simplest example is a horizontally oriented linear polarizer
for which $p_H = 1$ and $p_V = 0$. In practice, however, all
values $0 \leq p_H \leq 1$ and $0 \leq p_V \leq 1$ are attainable
(see, e.g.,~\cite{mataloni-2010}), which leads to a two-parameter
family of qubit-to-qutrit channels
\begin{eqnarray}
&& \!\!\!\!\! \Gamma \left[ \left(%
\begin{array}{cc}
  \varrho_{HH} & \varrho_{HV} \\
  \varrho_{VH} & \varrho_{VV} \\
\end{array}%
\right) \right] \nonumber\\
&& \!\!\!\!\! = \left(%
\begin{array}{cc|c}
  p_H \varrho_{HH} & \sqrt{p_H p_V} \varrho_{HV} & 0 \\
  \sqrt{p_H p_V} \varrho_{VH} & p_V \varrho_{VV} & 0 \\
  \hline
  0 & 0 & \begin{array}{c}
    (1-p_H)\varrho_{HH} \\
    + (1-p_V)\varrho_{VV} \\
  \end{array}  \\
\end{array}%
\right), \quad \label{pdl-channel}
\end{eqnarray}

\noindent with $p_H$ and $p_V$ being the parameters. The extra
(third) dimension in Eq.~\eqref{pdl-channel} corresponds to the
vacuum contribution $\ket{{\rm vac}}$ that leads to no detector
clicks. If $p_H = p_V$, then we get the standard erasure
channel~\cite{grassl-1997,bennett-1997}. If $p_H \neq p_V$, then
Eq.~\eqref{pdl-channel} defines a generalized erasure
channel~\cite{filippov-2021} (cf. a similar but different concept
in Ref.~\cite{siddhu-2021}) induced by the trace decreasing
operation $\varrho \rightarrow \Lambda_F [\varrho] := F \varrho
F^{\dag}$, where
\begin{equation*}
F = \sqrt{p_H} \ket{H}\bra{H} + \sqrt{p_V} \ket{V}\bra{V},
\end{equation*}

\noindent $\ket{H}$ and $\ket{V}$ are the single-photon states
with horizontal and vertical polarization, respectively. The brief
version of Eq.~\eqref{pdl-channel} is
\begin{equation*}
\Gamma[\varrho] = F \varrho F^{\dag} \oplus {\rm tr}[(I -
F^{\dag}F)\varrho] \, \ket{{\rm vac}}\bra{{\rm vac}}.
\end{equation*}

\noindent The term ${\rm tr}[(I - F^{\dag}F)\varrho]$ is the state
dependent erasure probability. Denoting $G := \sqrt{I -
F^{\dag}F}$ and recalling the notation $\Lambda_G [\varrho] := G
\varrho G^{\dag}$, the channel \eqref{pdl-channel} takes the form
\begin{equation} \label{Gamma-through-Lambda-and-Tr}
\Gamma = \Lambda_F \oplus ({\rm Tr} \circ \Lambda_G),
\end{equation}

\noindent where ${\rm Tr}$ denotes the trash-and-prepare map
$\varrho \rightarrow {\rm tr}[\varrho] \, \ket{{\rm vac}}\bra{{\rm
vac}}$. Interestingly, a complementary channel
$\widetilde{\Gamma}$ can be expressed as~\cite{filippov-2021}
\begin{equation*}
\widetilde{\Gamma} = \Lambda_G \oplus ({\rm Tr} \circ \Lambda_F),
\end{equation*}

\noindent which is equivalent to the change $p_H \rightarrow 1-
p_H$ and $p_V \rightarrow 1- p_V$ in Eq.~\eqref{pdl-channel}.

The fact that $\Gamma$ and $\widetilde{\Gamma}$ have the same
structure was used in Ref.~\cite{filippov-2021} to prove that
$\Gamma$ is antidegradable [so that $Q(\Gamma)=0$] if and only if
$\max(p_H,p_V) \leq \frac{1}{2}$ or $p_H=0$ or $p_V=0$, see the
green (medium gray) region in Fig.~\ref{figure-1}. It was also
shown in Ref.~\cite{filippov-2021} that $Q(\Gamma) > 0$ beyond the
antidegradability region. $\Gamma$ is degradable [so that
$Q(\Gamma)=Q_1(\Gamma)$] if and only if $\min(p_H,p_V) \geq
\frac{1}{2}$ or $p_H = 1$ or $p_V = 1$, see the red (dark gray)
region in Fig.~\ref{figure-1}. The final result of
Ref.~\cite{filippov-2021} is the analytical proof of
superadditivity relation $Q_2(\Gamma)
> Q_1(\Gamma)$ for two regions of attenuation factors: (i) $\frac{1}{2} < p_H < 1$ and
$0 < p_V < 1 - p_H$, (ii) $\frac{1}{2} < p_V < 1$ and $ 0 < p_H <
1 - p_V$; see the yellow (light gray) areas in
Fig.~\ref{figure-1}. White regions in Fig.~\ref{figure-1} are
\textit{terra incognita}, where neither the degradability nor the
antidegradability holds, and no strategies are known to outperform
the one-shot capacity $Q_1(\Gamma)$.

The goal of this paper is twofold. First, we are going to close
the gap in our understanding of the coherent-information
superadditivity region in Fig.~\ref{figure-1}. To do so we provide
some physically motivated $n$-partite entangled states
$\varrho^{(n)}$, using which the coherent-information
superadditivity region extends further and completely covers the
white area in Fig.~\ref{figure-1} in the limit $n \rightarrow
\infty$. This result is interesting per se as it presents an
analytical proof of the coherent-information superadditivity for
an arbitrary $n \geq 2$. Second, for fixed values of $p_H$ and
$p_V$, we are interested in finding particular states
$\varrho^{(n)}$ leading to higher values of the coherent
information. In this regard, we propose a scheme enabling one to
get a higher quantum communication rate by doubling the number of
channel uses.


\section{Superadditivity identification} \label{section-superadditivity-identification}

Technically, it is quite difficult to maximize the coherent
information $I_c(\varrho^{(n)},\Gamma^{\otimes n})$ with respect
to $n$-qubit density operators $\varrho^{(n)}$ even if $p_H$,
$p_V$, and $n$ are all fixed. In the case of the one-shot capacity
($n=1$), the optimal state $\varrho_{\rm opt}^{(1)}$ is shown to
be diagonal in the basis $\ket{H}$, $\ket{V}$ for all $p_H$ and
$p_V$, i.e.,
\begin{equation*}
\varrho_{\rm opt}^{(1)} = \varrho_{HH} \ket{H}\bra{H} +
\varrho_{VV} \ket{V}\bra{V};
\end{equation*}

\noindent however, a closed-form expression for the coefficients
$\varrho_{HH}$ and $\varrho_{VV}$ is still missing so they appear
as a solution of some equation that can be readily solved
numerically~\cite{filippov-2021}. If $\Gamma$ is not
antidegradable, then both $\varrho_{HH}
> 0$ and $\varrho_{VV}
> 0$ so that $Q_1(\Gamma) = I_c(\varrho_{\rm opt}^{(1)},\Gamma) >
0$. Therefore, a random subspace code to attain $Q_1(\Gamma)$ does
not need to exploit superpositions of horizontally and vertically
polarized photons in its ensemble states. If the degradability
property holds for $\Gamma$ [see the red (dark gray) region in
Fig.~\ref{figure-1}], then $Q(\Gamma) = Q_1(\Gamma) = \frac{1}{n}
I_c\left( (\varrho_{\rm opt}^{(1)}))^{\otimes n},\Gamma^{\otimes
n} \right)$ and there is no need nor benefit to consider states
$\varrho^{(n)}$ other than $(\varrho_{\rm opt}^{(1)}))^{\otimes
n}$. If $\Gamma$ is neither degradable nor antidegradable, then
there is a potential for improvement. In
Section~\ref{section-N-equals-2}, we review in detail an approach
of Ref.~\cite{filippov-2021} to find a two-qubit state
$\varrho^{(2)}$ outperforming $(\varrho_{\rm opt}^{(1)}))^{\otimes
2}$ in value of the two-shot coherent information
$I_c(\varrho,\Gamma^{\otimes 2})$ for some parameters $p_H$ and
$p_V$. In Section~\ref{section-N-greater-than-2}, we generalize
that approach to lower bound the $n$-shot quantum capacity
$Q_n(\Gamma)$ for an arbitrary number $n$ of channel uses.

\subsection{Two-shot capacity} \label{section-N-equals-2}

Suppose $n=2$. Consider the state
\begin{eqnarray} \label{rho-2}
\varrho^{(2)} &=& \big( \varrho_{\rm opt}^{(1)} \big)^{\otimes 2}
+ \varrho_{HH}\varrho_{VV} \left( \ket{HV}\bra{VH} +
\ket{VH}\bra{HV} \right) \nonumber\\
&=& \varrho_{HH}^2 \ket{HH}\bra{HH} + \varrho_{VV}^2
\ket{VV}\bra{VV} \nonumber\\
&& + 2 \varrho_{HH}\varrho_{VV} \frac{\ket{HV}+\ket{VH}}{\sqrt{2}}
\, \frac{\bra{HV}+\bra{VH}}{\sqrt{2}}.
\end{eqnarray}

\noindent Clearly, the diagonals of density matrices
$\varrho^{(2)}$ and $( \varrho_{\rm opt}^{(1)} )^{\otimes 2}$
coincide in the standard basis
$(\ket{HH},\ket{HV},\ket{VH},\ket{VV})$. The two photon states
$\ket{HV}$ and $\ket{VH}$ experience the same attenuation even if
$p_H \neq p_V$ due to the obvious symmetry. In fact, all vectors
from the subspace ${\cal H}_{1,1}:={\rm Span}(\ket{HV},\ket{VH})$
are equally attenuated, which makes it easy to calculate the
output state
\begin{eqnarray*}
&& \!\!\!\!\! \Gamma^{\otimes 2} [\varrho^{(2)}] \nonumber\\
&& \!\!\!\!\!  = \big( \Gamma[\varrho_{\rm opt}^{(1)}]
\big)^{\otimes 2} + p_H p_V \varrho_{HH}\varrho_{VV} \left(
\ket{HV}\bra{VH} +
\ket{VH}\bra{HV} \right) \nonumber\\
&& \!\!\!\!\!  = p_H^2 \varrho_{HH}^2 \ket{HH}\bra{HH} + p_V^2
\varrho_{VV}^2
\ket{VV}\bra{VV} \nonumber\\
&& \!\!\!\!\!  + 2 p_H p_V \varrho_{HH}\varrho_{VV}
\frac{\ket{HV}+\ket{VH}}{\sqrt{2}} \,
\frac{\bra{HV}+\bra{VH}}{\sqrt{2}} \nonumber\\
&& \!\!\!\!\!  + \big( (1-p_H)\varrho_{HH} + (1-p_V)\varrho_{VV}
\big)
\nonumber\\
&& \quad \times \left( \varrho_{\rm opt}^{(1)} \otimes \ket{{\rm
vac}}\bra{{\rm vac}} + \ket{{\rm vac}}\bra{{\rm vac}} \otimes
\varrho_{\rm opt}^{(1)}
\right) \nonumber\\
&& \!\!\!\!\!  + \big( (1-p_H)\varrho_{HH} + (1-p_V)\varrho_{VV}
\big)^2 \ket{{\rm vac}}\bra{{\rm vac}} \otimes \ket{{\rm
vac}}\bra{{\rm vac}}.
\end{eqnarray*}

\noindent The density operators $\Gamma^{\otimes 2}
[\varrho^{(2)}]$ and $\big( \Gamma[\varrho_{\rm opt}^{(1)}]
\big)^{\otimes 2}$ differ by their action in the subspace ${\cal
H}_{1,1}$, namely, $\Gamma^{\otimes 2} [\varrho^{(2)}]$ acts as a
coherent operator
\begin{equation} \label{different-block-1}
2 p_H p_V \varrho_{HH}\varrho_{VV}
\frac{\ket{HV}+\ket{VH}}{\sqrt{2}} \,
\frac{\bra{HV}+\bra{VH}}{\sqrt{2}},
\end{equation}

\noindent whereas $\big( \Gamma[\varrho_{\rm opt}^{(1)}]
\big)^{\otimes 2}$ acts as an incoherent operator
\begin{equation} \label{different-block-2}
p_H p_V \varrho_{HH}\varrho_{VV} ( \ket{HV}\bra{HV} +
\ket{VH}\bra{VH} ).
\end{equation}

\noindent This leads to a readily accountable difference in
spectra of the two states. Spectrum of~\eqref{different-block-1}
is $(2 p_H p_V \varrho_{HH}\varrho_{VV},0)$ and that
of~\eqref{different-block-2} is $(p_H p_V
\varrho_{HH}\varrho_{VV},p_H p_V \varrho_{HH}\varrho_{VV})$. We
have
\begin{eqnarray*}
S(\Gamma^{\otimes 2} [\varrho^{(2)}]) &=& S\Big( (
\Gamma[\varrho_{\rm opt}^{(1)}] )^{\otimes 2} \Big) \nonumber\\
&& - (2 \log 2) p_H p_V \varrho_{HH}\varrho_{VV}.
\end{eqnarray*}

\noindent As the complementary channel $\widetilde{\Gamma}$ is
obtained from the direct channel $\Gamma$ by the change $p_H
\rightarrow 1- p_H$ and $p_V \rightarrow 1- p_V$, we readily have
\begin{eqnarray*}
S(\widetilde{\Gamma}^{\otimes 2} [\varrho^{(2)}]) & = & S \Big( (
\widetilde{\Gamma}[\varrho_{\rm opt}^{(1)}] )^{\otimes 2} \Big)
\nonumber\\
&& - (2 \log 2) (1-p_H) (1-p_V) \varrho_{HH}\varrho_{VV}.
\end{eqnarray*}

\noindent Finally, we get
\begin{eqnarray}
&& I_c(\varrho^{(2)},\Gamma^{\otimes 2}) = S(\Gamma^{\otimes 2}
[\varrho^{(2)}]) - S(\widetilde{\Gamma}^{\otimes 2}
[\varrho^{(2)}])
\nonumber\\
&& = S\Big( \big( \Gamma[\varrho_{\rm opt}^{(1)}] \big)^{\otimes
2} \Big) - S \Big( \big( \widetilde{\Gamma}[\varrho_{\rm
opt}^{(1)}] \big)^{\otimes 2}
\Big) \nonumber\\
&& \quad + (2 \log 2) (1 - p_H - p_V)
\varrho_{HH}\varrho_{VV} \nonumber\\
&& = 2 I_c(\varrho_{\rm opt}^{(1)},\Gamma) + (2 \log 2) (1 - p_H -
p_V) \varrho_{HH}\varrho_{VV} \nonumber\\
&& = 2 Q_1(\Gamma) + (2 \log 2) (1 - p_H - p_V)
\varrho_{HH}\varrho_{VV}. \label{I-c-varrho-2}
\end{eqnarray}

\noindent The coherent information is superadditive if $(1 - p_H -
p_V) \varrho_{HH}\varrho_{VV} > 0$, i.e., if $p_H + p_V < 1$ and
the state $\varrho_{\rm opt}^{(1)}$ is nondegenerate. The latter
condition is fulfilled if $\Gamma$ is not antidegradable.
Combining these conditions we get two yellow (light gray) regions
in Fig.~\ref{figure-1}, where
\begin{equation*}
Q_2(\Gamma) \geq \frac{1}{2} I_c(\varrho^{(2)},\Gamma^{\otimes 2})
> Q_1(\Gamma).
\end{equation*}

\subsection{$n$-shot capacity} \label{section-N-greater-than-2}

Suppose $n > 2$. A generalization of the approach in
Section~\ref{section-N-equals-2} would be to consider a state $(
\varrho_{\rm opt}^{(1)} )^{\otimes n}$ and modify it to a state
$\varrho^{(n)}$, which would differ from $( \varrho_{\rm
opt}^{(1)} )^{\otimes n}$ when acting on some subspace that is
symmetric with respect to permutations of photons. Physically, the
subspace is to be chosen in such a way as to ensure a high enough
detection probability for all states from the subspace. Suppose
$p_H > p_V$, then the state $\ket{H}^{\otimes n}$ has the highest
detection probability, but the corresponding subspace ${\cal
H}_{n,0} := {\rm Span}(\ket{H}^{\otimes n})$ is trivial (has
dimension $1$). So we consider the subspace ${\cal H}_{n-1,1}$
spanned by the vector $\ket{H}^{\otimes (n-1)} \otimes \ket{V}$
and all its photon-permuted versions. The detection probability
for all states from this subspace equals $p_H^{n-1} p_V$. The
following entangled $n$-qubit $W$-state belongs to ${\cal
H}_{n-1,1}$:
\begin{eqnarray} \label{W-state}
\ket{W^{(n)}} =  \frac{1}{\sqrt{n}} & \Big( & \ket{\underbrace{H H
\ldots H H}_{n-1}
V} \nonumber\\
&+& \ket{\underbrace{H H \ldots H}_{n-2} V H} \nonumber\\
&+& \ldots \nonumber\\
&+& \ket{V \underbrace{H \ldots H H H}_{n-1}} \Big) \in {\cal
H}_{n-1,1}.
\end{eqnarray}

Consider the $n$-qubit density operator $\varrho^{(n)}$ defined
through
\begin{equation*}
\varrho^{(n)} \ket{\varphi} = \left\{
\begin{array}{ll}
  (\varrho_{\rm opt}^{(1)})^{\otimes n} \ket{\varphi} & \text{if~} \ket{\varphi} \perp {\cal H}_{n-1,1}, \\
  n \varrho_{HH}^{n-1} \varrho_{VV} \ket{W^{(n)}} \bra{W^{(n)}} \ket{\varphi} & \text{if~} \ket{\varphi} \in {\cal H}_{n-1,1}. \\
\end{array} \right.
\end{equation*}

\noindent The restriction of $\varrho^{(n)}$ to the subspace
${\cal H}_{n-1,1}$ is a coherent (rank-1) operator
\begin{equation} \label{coh-part}
\lfloor \varrho^{(n)} \rfloor_{{\cal H}_{n-1,n}} := n
\varrho_{HH}^{n-1} \varrho_{VV} \ket{W^{(n)}}\bra{W^{(n)}},
\end{equation}

\noindent whereas the restriction of $(\varrho_{\rm
opt}^{(1)})^{\otimes n}$ to the subspace ${\cal H}_{n-1,1}$ is a
mixed (rank-$n$) operator
\begin{eqnarray}
&& \lfloor (\varrho_{\rm opt}^{(1)})^{\otimes n} \rfloor_{{\cal
H}_{n-1,n}} \nonumber\\
&& :=  \varrho_{HH}^{n-1} \varrho_{VV} \Big( \ket{ \underbrace{H H
\ldots H H}_{n-1} V}\bra{\underbrace{H H
\ldots H H}_{n-1} V} \nonumber\\
&& \qquad\qquad\qquad +  \ket{\underbrace{H H \ldots H}_{n-2} V H}\bra{\underbrace{H H \ldots H}_{n-2} V H} \nonumber\\
&& \qquad\qquad\qquad +  \ldots \nonumber\\
&& \qquad\qquad\qquad +  \ket{V \underbrace{H \ldots H H
H}_{n-1}}\bra{V \underbrace{H \ldots H H H}_{n-1}} \Big), \qquad
\label{incoh-part}
\end{eqnarray}

\noindent but beyond that restriction
\begin{equation*}
\varrho^{(n)} - \lfloor \varrho^{(n)} \rfloor_{{\cal H}_{n-1,n}} =
(\varrho_{\rm opt}^{(1)})^{\otimes n} - \lfloor (\varrho_{\rm
opt}^{(1)})^{\otimes n} \rfloor_{{\cal H}_{n-1,n}}.
\end{equation*}

Using the direct sum
representation~\eqref{Gamma-through-Lambda-and-Tr} of the channel
$\Gamma$, we explicitly find its tensor power
\begin{eqnarray}
\Gamma^{\otimes n} &=& \Lambda_F^{\otimes n} \oplus \ldots
\nonumber\\
&& \oplus \underbrace{ \left[\Lambda_F^{\otimes(n-k)} \otimes
({\rm Tr} \circ \Lambda_G)^{\otimes k} \right] \oplus \ldots}_{\binom{n}{k} \text{~terms}} \nonumber\\
&& \oplus \ldots \oplus ({\rm Tr} \circ \Lambda_G)^{\otimes n},
\label{Gamma-n}
\end{eqnarray}

\noindent where the brace denotes a direct sum of $\binom{n}{k}$
different terms, with each term being a permuted tensor product of
$n-k$ maps $\Lambda_F$ and $k$ maps ${\rm Tr} \circ \Lambda_G$.
Let us consider how the term $\Lambda_F^{\otimes(n-k)} \otimes
({\rm Tr} \circ \Lambda_G)^{\otimes k}$ affects the operators
$\lfloor \varrho^{(n)} \rfloor_{{\cal H}_{n-1,n}}$ and $\lfloor
(\varrho_{\rm opt}^{(1)})^{\otimes n} \rfloor_{{\cal H}_{n-1,n}}$.
Recalling the effect of the partial trace on $W$-states, we see
that the coherent component of $\Lambda_F^{\otimes(n-k)} \otimes
({\rm Tr} \circ \Lambda_G)^{\otimes k} \big[ \lfloor \varrho^{(n)}
\rfloor_{{\cal H}_{n-1,n}} \big]$ reads
\begin{eqnarray}
&& \varrho_{HH}^{n-1} \varrho_{VV} \, p_H^{n-k-1} p_V (1-p_H)^k
\nonumber\\
&& \times (n-k)\ket{W^{(n-k)}}\bra{W^{(n-k)}} \otimes (\ket{{\rm
vac}}\bra{{\rm vac}})^{\otimes k}, \qquad \label{coh-component}
\end{eqnarray}

\noindent whereas $\Lambda_F^{\otimes(n-k)} \otimes ({\rm Tr}
\circ \Lambda_G)^{\otimes k} \big[ \lfloor (\varrho_{\rm
opt}^{(1)})^{\otimes n} \rfloor_{{\cal H}_{n-1,n}} \big]$ has the
completely incoherent component
\begin{eqnarray}
&& \varrho_{HH}^{n-1} \varrho_{VV} \, p_H^{n-k-1} p_V (1-p_H)^k
\nonumber\\
&& \times  \Big( \ket{ \underbrace{H H \ldots H H}_{n-k-1}
V}\bra{\underbrace{H H
\ldots H H}_{n-k-1} V} \nonumber\\
&& \qquad +  \ket{\underbrace{H H \ldots H}_{n-k-2} V H}\bra{\underbrace{H H \ldots H}_{n-k-2} V H} \nonumber\\
&& \qquad +  \ldots \nonumber\\
&& \qquad +  \ket{V \underbrace{H \ldots H H H}_{n-k-1}}\bra{V
\underbrace{H \ldots H H H}_{n-k-1}} \Big) \nonumber\\
&& \otimes (\ket{{\rm vac}}\bra{{\rm vac}})^{\otimes k}.
\label{incoh-component}
\end{eqnarray}

\noindent The operator~\eqref{coh-component} has the only nonzero
eigenvalue, whereas the operator~\eqref{incoh-component} has $n-k$
coincident nonzero eigenvalues, with traces of the two operators
being the same. Therefore, the only nonzero eigenvalue of the
operator~\eqref{coh-component} is $(n-k)$ multiplied by any
nonzero eigenvalue of the operator~\eqref{incoh-component}. This
leads to a simple expression for the difference in entropies,
namely,
\begin{eqnarray} \label{entropy-decrement-term}
&& \!\!\!\!\! S \Big(\Lambda_F^{\otimes(n-k)} \otimes ({\rm Tr}
\circ
\Lambda_G)^{\otimes k} [\varrho^{(n)}] \Big) \nonumber\\
&& \!\!\!\!\! = S\Big( \Lambda_F^{\otimes(n-k)} \otimes ({\rm Tr}
\circ \Lambda_G)^{\otimes k} [(\varrho_{\rm opt}^{(1)})^{\otimes
n}]
\Big) \nonumber\\
&& \!\!\!\!\! \quad - \varrho_{HH}^{n-1} \varrho_{VV} \, p_V
p_H^{n-k-1} (1-p_H)^k \,(n-k) \log (n-k). \qquad
\end{eqnarray}

\noindent Since the operators~\eqref{coh-part}
and~\eqref{incoh-part} are invariant with respect to permutations
of photons, each term in the brace in Eq.~\eqref{Gamma-n} results
in the same entropy decrement as in
Eq.~\eqref{entropy-decrement-term}. Summing all the decrements, we
get
\begin{eqnarray*}
&& \!\!\!\!\!  S \Big( \Gamma^{\otimes n} [\varrho^{(n)}] \Big) =
S\Big( \Gamma^{\otimes n} [(\varrho_{\rm opt}^{(1)})^{\otimes n}]
\Big) - \varrho_{HH}^{n-1} \varrho_{VV}
\nonumber\\
&& \!\!\!\!\! \times \sum_{k=0}^{n-1} \binom{n}{k} \, p_V
p_H^{n-k-1} (1-p_H)^k \,(n-k) \log (n-k).
\end{eqnarray*}

\noindent Similarly, for the complementary channel we
have
\begin{eqnarray*}
&& \!\!\!\!\! S \Big( \widetilde{\Gamma}^{\otimes n}
[\varrho^{(n)}] \Big) = S\Big( \widetilde{\Gamma}^{\otimes n}
[(\varrho_{\rm
opt}^{(1)})^{\otimes n}] \Big) - \varrho_{HH}^{n-1} \varrho_{VV} \nonumber\\
&& \!\!\!\!\! \times \sum_{k=0}^{n-1} \binom{n}{k} \, (1-p_V)
(1-p_H)^{n-k-1} p_H^k \,(n-k) \log (n-k).
\end{eqnarray*}

\noindent Finally, we get
\begin{eqnarray}
&&  \!\!\!\!\!\!\!\!\!\! Q_n(\Gamma) - Q_1(\Gamma) \geq
\frac{1}{n} \left[ I_c(\varrho^{(n)},\Gamma^{\otimes n}) - I_c
\big( (\varrho_{\rm opt}^{(1)})^{\otimes n},\Gamma^{\otimes n}
\big) \right]
\nonumber\\
&&  \!\!\!\!\!\!\!\!\!\! = \frac{1}{n} \varrho_{HH}^{n-1}
\varrho_{VV} \sum_{k=0}^{n-1}
\binom{n}{k} (n-k) \log (n-k) \nonumber\\
&&  \!\!\!\!\!\!\!\!\!\! \quad \times \left[ (1-p_V)
(1-p_H)^{n-k-1} p_H^k
- p_V p_H^{n-k-1} (1-p_H)^k  \right] \nonumber\\
&&  \!\!\!\!\!\!\!\!\!\! =  \varrho_{HH}^{n-1} \varrho_{VV}
\sum_{k=0}^{n-1}
\binom{n-1}{k} \log (n-k) \nonumber\\
&&  \!\!\!\!\!\!\!\!\!\! \quad \times  \left[ (1-p_V)
(1-p_H)^{n-k-1} p_H^k
- p_V p_H^{n-k-1} (1-p_H)^k  \right] \nonumber\\
&&  \!\!\!\!\!\!\!\!\!\! =  \varrho_{HH}^{n-1} \varrho_{VV}
\sum_{k=0}^{n-1}
\binom{n-1}{k} (1-p_H)^{n-k-1} p_H^k \nonumber\\
&&  \!\!\!\!\!\!\!\!\!\! \quad \times  \left[ (1-p_V) \log (n-k)
- p_V \log (k+1) \right]. \label{n-benefit}
\end{eqnarray}

\noindent If the obtained expression~\eqref{n-benefit} is
positive, then we successfully identify the coherent-information
superadditivity in the form $Q_n(\Gamma) > Q_1(\Gamma)$. Suppose
$\Gamma$ is not antidegradable, then $\varrho_{HH}
> 0$, $\varrho_{VV} > 0$, and $Q_n(\Gamma) > Q_1(\Gamma)$ if the sum in Eq.~\eqref{n-benefit} is positive.

In the above analysis, we assumed $p_H > p_V$. The converse case
$p_V > p_H$ obviously reduces to the considered one if we replace
$\ket{H} \leftrightarrow \ket{V}$ in Eq.~\eqref{W-state}.
Therefore, we make the following conclusion: $Q_n(\Gamma) >
Q_1(\Gamma)$ if $\Gamma$ is not antidegradable and $w_n(p_H,p_V) >
0$, where
\begin{eqnarray}
&& w_n(p_H,p_V) := \sum\limits_{k=0}^{n-1} \binom{n-1}{k}
(1-p_H)^{n-k-1} p_H^k \nonumber\\
&& \qquad \times \left[ (1-p_V) \log (n-k)  - p_V \log (k+1)
\right] \label{w-n-1}
\end{eqnarray}

\noindent if $p_H > p_V$,
\begin{eqnarray}
&& w_n(p_H,p_V) := \sum\limits_{k=0}^{n-1} \binom{n-1}{k}
(1-p_V)^{n-k-1} p_V^k \nonumber\\
&& \qquad \times \left[ (1-p_H) \log (n-k)  - p_H \log (k+1)
\right] \label{w-n-2}
\end{eqnarray}

\noindent if $p_V > p_H$.

\begin{figure}
\includegraphics[width=8cm]{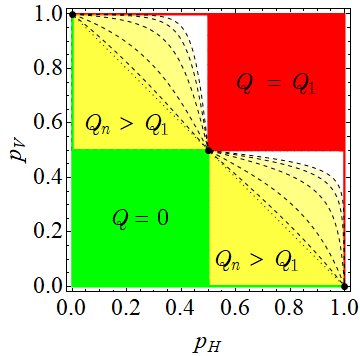}
\caption{\label{figure-2} Nested superadditivity regions
$Q_n(\Gamma)
> Q_1(\Gamma)$ [yellow (light gray) areas] in the parameter space
(dimensionless attenuation factors $p_H$ and $p_V$ for
horizontally and vertically polarized photons) with boundaries
depicted for various values: $n=2$ (dotted line),
$n=3,10,10^2,10^3,10^4$ (dashed lines from left to right).}
\end{figure}

In the case $n=2$, the condition $w_2(p_H,p_V) > 0$ is equivalent
to $p_H + p_V < 1$, i.e., we reproduce the results of
Section~\ref{section-N-equals-2}. If $n \geq 3$, then the region
of parameters $p_H$ and $p_V$, where $Q_n(\Gamma) > Q_1(\Gamma)$,
is strictly larger than the region, where $Q_2(\Gamma) >
Q_1(\Gamma)$, see Fig.~\ref{figure-2}. Interestingly, the greater
$n$ the larger the region, where $Q_n(\Gamma) > Q_1(\Gamma)$. If
$n=10^4$, then the condition $w_{10^4}(p_H,p_V) > 0$ defines a
region in the plane $(p_H,p_V)$, which almost coincides with the
area where $\Gamma$ is neither degradable nor antidegradable (see
Fig.~\ref{figure-2}). This observation motivates us to study the
asymptotic behaviour of $w_n(p_H,p_V)$.

The binomial distribution $\{\binom{n-1}{k} (1-p)^{n-k-1}
p^k\}_{k=0}^{n-1}$ tends to the normal distribution ${\cal
N}(np,np(1-p))$ with the mean value $np$ and the standard
deviation $\sqrt{np(1-p)}$ when $0<p<1$ and $n$ tends to
infinity~\cite{govindarajulu-1965}. Therefore, the terms with $k
\approx n p_H$ contribute the most to Eq.~\eqref{w-n-1} and the
terms with $k \approx n p_V$ contribute the most to
Eq.~\eqref{w-n-2}. In the asymptotic limit $n \rightarrow \infty$
we have
\begin{eqnarray}
w_n(p_H,p_V) & \approx & (1- 2p_V) \log n + (1-p_V) \log (1 - p_H)
\nonumber\\
&& - p_V \log p_H  \text{~if~} p_H > p_V, \label{w-n-asymptotic-1} \\
w_n(p_H,p_V) & \approx & (1- 2p_H) \log n + (1-p_H) \log (1 - p_V)
\nonumber\\
&& - p_H \log p_V  \text{~if~} p_V > p_H. \label{w-n-asymptotic-2}
\end{eqnarray}

\noindent Hence, $w_n(p_H,p_V) > 0$ in the asymptotic limit $n
\rightarrow \infty$ if $0 < p_V < \frac{1}{2} < p_H < 1$ or $0 <
p_H < \frac{1}{2} < p_V < 1$, which is exactly the region, where
$\Gamma$ is neither degradable nor antidegradable (see
Fig.~\ref{figure-2}).

Suppose the parameters $p_H$ and $p_V$ are fixed. Exploiting the
asymptotic formulas
\eqref{w-n-asymptotic-1}--\eqref{w-n-asymptotic-2} and solving the
inequality $w_n(p_H,p_V) \geq 0$, we estimate the number $n$
needed to observe the superadditivity phenomenon $Q_n > Q_1$:
\begin{equation*}
n \gtrsim n_0 := \left\{ \begin{array}{ll}
  \left(\frac{p_H^{p_V}}{(1 - p_H)^{1-p_V}}\right)^{\frac{1}{1-2p_V}} & \text{if~} 0 < p_V < \frac{1}{2} < p_H < 1, \\
  \left(\frac{p_V^{p_H}}{(1 - p_V)^{1-p_H}}\right)^{\frac{1}{1-2p_H}} & \text{if~} 0 < p_H < \frac{1}{2} <
p_V < 1. \\
\end{array} \right.
\end{equation*}

\noindent If $n \gg n_0$, then the proposed states yields the
following benefit in the quantum communication rate:
\begin{eqnarray*}
&& \frac{1}{n} I_c(\varrho^{(n)},\Gamma^{\otimes n}) \approx
Q_1(\Gamma) \nonumber\\
&& + \left\{
\begin{array}{ll}
   (1- 2p_V) \varrho_{HH}^{n-1} \varrho_{VV} \log n & \text{if~} 0 < p_V < \frac{1}{2} < p_H < 1, \\
  (1- 2p_H) \varrho_{HH} \varrho_{VV}^{n-1} \log n & \text{if~} 0 < p_H < \frac{1}{2} <
p_V < 1. \\
\end{array} \right.
\end{eqnarray*}


\section{Superadditivity improvement} \label{section-improvement}

The goal of the previous section was to detect the
coherent-information superadditivity in the widest region of
parameters $p_H$ and $p_V$. In this section we discuss how to get
a higher quantum communication rate (for fixed values of $p_H$ and
$p_V$) by using the channel multiple times.

Our approach is to combine two $n$-qubit states $\varrho^{(n)}$
from Section~\ref{section-superadditivity-identification} and
slightly modify them to get a better $2n$-qubit state
$\xi^{(2n)}$. To illustrate this approach, consider the region $0
< p_V < 1-p_H < \frac{1}{2}$, where $Q_2 > Q_1$ (see
Section~\ref{section-N-equals-2}). Let $\varrho^{(2)}$ be a
partially coherent state given by Eq.~\eqref{rho-2}. The
four-qubit state $\varrho^{(2)} \otimes \varrho^{(2)}$ inherits
some superpositions in the subspace spanned by 12 vectors:
$\ket{HHHV}$, $\ket{HHVH}$, $\ket{HVHH}$, $\ket{HVHV}$,
$\ket{HVVH}$, $\ket{HVVV}$, $\ket{VHHH}$, $\ket{VHHV}$,
$\ket{VHVH}$, $\ket{VHVV}$, $\ket{VVHV}$, and $\ket{VVVH}$. On the
other hand, the states $\ket{HHVV}$ and $\ket{VVHH}$ incoherently
contribute to $\varrho^{(2)} \otimes \varrho^{(2)}$ though they
have the same detection probability $p_H^2 p_V^2$. We use the
latter fact to construct a more coherent version of the state
$\varrho^{(2)} \otimes \varrho^{(2)}$ as follows:
\begin{eqnarray*}
\xi^{(4)} & := & \varrho^{(2)} \otimes \varrho^{(2)} +
\varrho_{HH}^2
\varrho_{VV}^2 ( \ket{HHVV}\bra{VVHH}  \nonumber\\
&& \qquad\qquad\qquad\qquad + \ket{VVHH}\bra{HHVV} ). \quad
\end{eqnarray*}

\noindent The states $\varrho^{(2)} \otimes \varrho^{(2)}$ and
$\xi^{(4)}$ have the almost identical spectra, with the difference
being in the eigenspace spanned by $\ket{HHVV}$ and $\ket{VVHH}$.
That difference is translated into the operators
$\Lambda_F^{\otimes 4}[\xi^{(4)}]$ and $\Lambda_F^{\otimes
4}[\varrho^{(2)} \otimes \varrho^{(2)}]$, which results in
\begin{equation*}
S(\Lambda_F^{\otimes 4}[\xi^{(4)}]) = S(\Lambda_F^{\otimes
4}[\varrho^{(2)} \otimes \varrho^{(2)}]) - (2 \log 2) p_H^2 p_V^2
\varrho_{HH}^2 \varrho_{VV}^2.
\end{equation*}

\noindent Since the partial trace of the operator $(
\ket{HHVV}\bra{VVHH} + \ket{VVHH}\bra{HHVV} )$ with respect to any
photon vanishes, this means that $\Lambda_F^{\otimes 3} \otimes
({\rm Tr} \circ \Lambda_G)[\xi^{(4)}] = \Lambda_F^{\otimes 3}
\otimes ({\rm Tr} \circ \Lambda_G)[\varrho^{(2)} \otimes
\varrho^{(2)}]$, etc., so that the density operators $\xi^{(4)}$
and $\varrho^{(2)} \otimes \varrho^{(2)}$ are both mapped to the
same operator when affected by any map involving the
trash-and-prepare operation ${\rm Tr}$ for at least one of the
qubits. Recalling the fact that $\Gamma^{\otimes 4} = [\Lambda_F
\oplus ({\rm Tr} \circ \Lambda_G) ]^{\otimes 4}$, we get
\begin{equation*}
S(\Gamma^{\otimes 4}[\xi^{(4)}]) = S(\Gamma^{\otimes
4}[\varrho^{(2)} \otimes \varrho^{(2)}]) - (2 \log 2) p_H^2 p_V^2
\varrho_{HH}^2 \varrho_{VV}^2.
\end{equation*}

\noindent Similarly,
\begin{eqnarray*}
S(\widetilde{\Gamma}^{\otimes 4}[\xi^{(4)}]) & = &
S(\widetilde{\Gamma}^{\otimes 4}[\varrho^{(2)} \otimes
\varrho^{(2)}]) \nonumber\\
&& - (2 \log 2) (1-p_H)^2 (1-p_V)^2 \varrho_{HH}^2 \varrho_{VV}^2.
\end{eqnarray*}

These relations lead to a greater coherent information as compared
to twice the expression~\eqref{I-c-varrho-2}, namely,
\begin{eqnarray}
&& I_c(\xi^{(4)},\Gamma^{\otimes 4}) = S(\Gamma^{\otimes
4}[\xi^{(4)}]) - S(\widetilde{\Gamma}^{\otimes 4}[\xi^{(4)}])
\nonumber\\
&& = S(\Gamma^{\otimes 4}[\varrho^{(2)} \otimes \varrho^{(2)}]) -
S(\widetilde{\Gamma}^{\otimes 4}[\varrho^{(2)} \otimes
\varrho^{(2)}]) \nonumber\\
&& \quad + (2 \log 2) \varrho_{HH}^2 \varrho_{VV}^2 [
(1-p_H)^2 (1-p_V)^2 - p_H^2 p_V^2] \nonumber\\
&& = 2 I_c(\varrho^{(2)},\Gamma^{\otimes 2}) \nonumber\\
&& \quad + (2 \log 2) \varrho_{HH}^2 \varrho_{VV}^2 [ (1-p_H)^2
(1-p_V)^2 - p_H^2 p_V^2]
\nonumber\\
&& = 4 Q_1(\Gamma) + (4 \log 2) \varrho_{HH}\varrho_{VV} (1 - p_H
- p_V) \nonumber\\
&& \quad + (2 \log 2) \varrho_{HH}^2 \varrho_{VV}^2 [ (1-p_H)^2
(1-p_V)^2 - p_H^2 p_V^2]. \qquad \label{I-c-rho-4}
\end{eqnarray}

\noindent Dividing Eq.~\eqref{I-c-rho-4} by 4, we get a better
lower bound
\begin{eqnarray}
&& Q_4(\Gamma) - Q_1(\Gamma) \geq \frac{1}{4}
I_c(\xi^{(4)},\Gamma^{\otimes
4}) - Q_1(\Gamma) \nonumber\\
&& = \left[
 1 + \frac{1}{2} \varrho_{HH}
\varrho_{VV}  ( 1 - p_H - p_V + 2 p_H p_V ) \right] \nonumber\\
&& \times (\log 2) (1 - p_H - p_V) \varrho_{HH}\varrho_{VV}.
\label{Q-4-modified}
\end{eqnarray}

\noindent The lower bound~\eqref{Q-4-modified} significantly
outperforms the lower bound~\eqref{n-benefit} for $n=4$ in a wide
range of parameters $p_H$ and $p_V$. For instance, if $p_H=0.7$
and $p_V=0.2$, then Eq.~\eqref{Q-4-modified} yields $Q_4(\Gamma) -
Q_1(\Gamma) \geq 6.3 \times 10^{-3}$ bits, whereas
Eq.~\eqref{n-benefit} yields $Q_4(\Gamma) - Q_1(\Gamma) \geq 9.1
\times 10^{-5}$ bits.

Clearly, the presented approach works well to extend the $n$-qubit
state $\varrho^{(n)}$ from Section~\ref{section-N-greater-than-2}
to a $2n$-qubit state $\xi^{(2n)}$ by modifying the state
$(\varrho^{(n)})^{\otimes 2}$ in the subspace spanned by
$\ket{H}^{\otimes n} \otimes \ket{V}^{\otimes n}$ and
$\ket{V}^{\otimes n} \otimes \ket{H}^{\otimes n}$. Similarly, the
modified $2n$-qubit state $\xi^{(2n)}$ can further be improved to
a $4n$-qubit state and so on ad infinitum. Starting with the
two-qubit state in Section~\ref{section-N-equals-2}, we get the
following result:
\begin{eqnarray*}
&& Q(\Gamma) - Q_1(\Gamma) \geq (\log 2)(1 - p_H - p_V)
\sum\limits_{m=0}^{\infty}
\frac{\varrho_{HH}^{2^m}\varrho_{VV}^{2^m}}{2^m} \nonumber\\
&& \qquad\qquad \times \sum\limits_{k=0}^{2^m-1} (1-p_H)^{2^m-k-1}
(1-p_V)^{2^m-k-1} p_H^k p_V^k.
\end{eqnarray*}


\section{Conclusions} \label{section-conclusions}

A phenomenon of the coherent-information superadditivity makes it
possible to enhance the quantum communication rate by using clever
codes. In this paper, we have studied the superadditivity
phenomenon in physically relevant quantum communication lines with
polarization dependent losses. Such lines represent a
two-parameter family of generalized erasure channels $\Gamma$,
with the attenuation factors $p_H$ and $p_V$ for horizontally and
vertically polarized photons being the parameters. In prior
research, two-shot capacity $Q_2(\Gamma)$ was shown to be greater
than the one-shot capacity $Q_1(\Gamma)$ for some values of $p_H$
and $p_V$ within the region $p_H+p_V<1$~\cite{filippov-2021}.
Interestingly, if $p_H+p_V \geq 1$, then $\Gamma$ is
input-degradable in the sense that there exists a quantum channel
$\Upsilon$ such that $\widetilde{\Gamma} = \Gamma \circ \Upsilon$.
Making an analogy with the case of standard degradable channels,
it is tempting to conjecture that the input-degradability implies
$Q_n(\Gamma) = Q_1(\Gamma)$ if $p_H+p_V \geq 1$. Our study shows
that this conjecture is false: the 3-qubit state $\varrho^{(3)}$
in Section~\ref{section-N-greater-than-2} insures $Q_3(\Gamma) >
Q_1(\Gamma)$ if $p_H+p_V = 1$ and $0 \neq p_H \neq p_V \neq 0$,
see Fig.~\ref{figure-2}.

The more the number of channel uses in
Section~\ref{section-N-greater-than-2} the wider the region of
parameters $p_H$ and $p_V$, where the superadditivity phenomenon
takes place. In the limit of infinitely many channel uses, we have
proved the strict inequality $Q(\Gamma)
> Q_1(\Gamma)$ for all $p_H$ and $p_V$ satisfying $0 < p_V <
\frac{1}{2} < p_H < 1$ or $0 < p_H < \frac{1}{2} < p_V < 1$, i.e.,
$Q(\Gamma) > Q_1(\Gamma)$ whenever $\Gamma$ is neither degradable
nor antidegradable. A feature of the state proposed in
Section~\ref{section-N-greater-than-2} is that it has a clear
physical meaning: $\varrho^{(n)}$ has an entangled component
proportional to $\ket{W^{(n)}}\bra{W^{(n)}}$, which in turn has a
high detection probability and whose structure is preserved by
polarization dependent losses due to the permutation symmetry.
Clearly, one could alternatively use another Dicke
state~\cite{dicke-1954,chen-2020} instead of $\ket{W^{(n)}}$;
however, the detection probability would be less in that case.

In this work, we were interested not only in the superadditivity
identification but also in its improvement with the increase of
channel uses. In Section~\ref{section-N-greater-than-2}, we
proposed a method how to get a higher quantum communication rate
by doubling the number of channel uses. We believe that the scheme
is far from being optimal, which necessitates a further search of
better codes, e.g., by using a neural network state
ansatz~\cite{bausch-leditzky,banik-2022}. Nonetheless, our
analytically derived states with known asymptotic values of
coherent information may serve as a benchmark for future codes
generated by numerical optimization.

\begin{acknowledgements}
The author thanks Maksim E. Shirokov for fruitful comments. This work was supported by the Russian Science Foundation under grant no. 19-11-00086, \href{https://rscf.ru/en/project/19-11-00086/}{https://rscf.ru/en/project/19-11-00086/}.
\end{acknowledgements}


\begin{thebibliography}{99}

\bibitem{horodecki-2009}
R. Horodecki, P. Horodecki, M. Horodecki, and K. Horodecki,
Quantum entanglement, Rev. Mod. Phys. {\bf 81}, 865 (2009).

\bibitem{erhard-2020}
M. Erhard, M. Krenn, and A. Zeilinger, Advances in
high-dimensional quantum entanglement, Nat. Rev. Phys. {\bf 2},
365 (2020).

\bibitem{lloyd-1997}
S. Lloyd, Capacity of the noisy quantum channel, Phys. Rev. A {\bf
55}, 1613 (1997).

\bibitem{barnum-1998}
H. Barnum, M. A. Nielsen, and B. Schumacher, Information
transmission through a noisy quantum channel, Phys. Rev. A {\bf
57}, 4153 (1998).

\bibitem{shor-2002}
P. W. Shor, Quantum error correction, Lecture notes of MSRI
Workshop on Quantum Information And Cryptography (November 4--8,
2002). Available at
https://www.msri.org/workshops/203/schedules/1181.

\bibitem{devetak-2005}
I. Devetak, The private classical capacity and quantum capacity of
a quantum channel, IEEE Transactions on Information Theory {\bf
51}, 44 (2005).

\bibitem{wilde-2013}
M. M. Wilde, {\it Quantum Information Theory} (Cambridge
University Press, Cambridge, 2013).

\bibitem{holevo-2012}
A. S. Holevo, {\it Quantum Systems, Channels, Information. A
Mathematical Introduction} (de Gruyter, Berlin, Boston, 2012).

\bibitem{devetak-shor-2005}
I. Devetak and P. Shor, The capacity of a quantum channel for
simultaneous transmission of classical and quantum information,
Commun. Math. Phys. {\bf 256}, 287 (2005).

\bibitem{divincenzo-1998}
D. P. DiVincenzo, P. W. Shor, and J. A. Smolin, Quantum-channel
capacity of very noisy channels, Phys. Rev. A {\bf 57}, 830
(1998).

\bibitem{fern-2008}
J. Fern and K. B. Whaley, Lower bounds on the nonzero capacity of
Pauli channels, Phys. Rev. A {\bf 78}, 062335 (2008).

\bibitem{leditzky-2018}
F. Leditzky, D. Leung, and G. Smith, Dephrasure channel and
superadditivity of coherent information, Phys. Rev. Lett. {\bf
121}, 160501 (2018).

\bibitem{yu-2020}
S. Yu, Y. Meng, R. B. Patel, Y.-T. Wang, Z.-J. Ke, W. Liu, Z.-P.
Li, Y.-Z. Yang, W.-H. Zhang, J.-S. Tang, C.-F. Li, and G.-C. Guo,
Experimental observation of coherent-information superadditivity
in a dephrasure channel, Phys. Rev. Lett. {\bf 125}, 060502
(2020).

\bibitem{siddhu-2021}
V. Siddhu and R. B. Griffiths, Positivity and nonadditivity of
quantum capacities using generalized erasure channels, IEEE Trans.
Inform. Theory {\bf 67}, 4533 (2021).

\bibitem{siddhu-nov-2020}
V. Siddhu, Leaking information to gain entanglement,
arXiv:2011.15116.

\bibitem{leditzky-2022}
F. Leditzky, D. Leung, V. Siddhu, G. Smith, J. A. Smolin, Generic
nonadditivity of quantum capacity in simple channels,
arXiv:2202.08377.

\bibitem{cubitt-2015}
T. Cubitt, D. Elkouss, W. Matthews, M. Ozols, D.
P\'{e}rez-Garc\'{\i}a, and S. Strelchuk, Unbounded number of
channel uses may be required to detect quantum capacity, Nature
Commun. {\bf 6}, 6739 (2015).

\bibitem{gisin-1997}
N. Gisin and B. Huttner, Combined effects of polarization mode
dispersion and polarization dependent losses in optical fibers,
Optics Communications {\bf 142}, 119 (1997).

\bibitem{kirby-2019}
B. T. Kirby, D. E. Jones, and M. Brodsky, Effect of polarization
dependent loss on the quality of transmitted polarization
entanglement, Journal of Lightwave Technology {\bf 37}, 95 (2019).

\bibitem{li-2018}
C. Li, M. Curty, F. Xu, O. Bedroya, and H.-K. Lo, Secure quantum
communication in the presence of phase- and polarization-dependent
loss, Phys. Rev. A {\bf 98}, 042324 (2018).

\bibitem{filippov-QP-2021}
S. N. Filippov, Trace decreasing quantum dynamical maps:
Divisibility and entanglement dynamics, arXiv:2108.13372.

\bibitem{filippov-quanta-2021}
S. N. Filippov, Entanglement robustness in trace decreasing
quantum dynamics, Quanta {\bf 10}, 15 (2021).

\bibitem{filippov-2021}
S. N. Filippov, Capacity of trace decreasing quantum operations
and superadditivity of coherent information for a generalized
erasure channel, J. Phys. A: Math. Theor. {\bf 54}, 255301 (2021).

\bibitem{mataloni-2010}
I. Bongioanni, L. Sansoni, F. Sciarrino, G. Vallone, and P.
Mataloni, Experimental quantum process tomography of
non-trace-preserving maps, Phys. Rev. A {\bf 82}, 042307 (2010).

\bibitem{grassl-1997}
M. Grassl, T. Beth, and T. Pellizzari, Codes for the quantum
erasure channel, Phys. Rev. A {\bf 56}, 33 (1997).

\bibitem{bennett-1997}
C. H. Bennett, D. P. DiVincenzo, and J. A. Smolin, Capacities of
quantum erasure channels, Phys. Rev. Lett. {\bf 78}, 3217 (1997).

\bibitem{govindarajulu-1965}
Z. Govindarajulu, Normal approximations to the classical discrete
distributions, Sankhy{\=a}: The Indian Journal of Statistics,
Series A {\bf 27}, 143 (1965).

\bibitem{dicke-1954}
R. H. Dicke, Coherence in spontaneous radiation processes, Phys.
Rev. {\bf 93}, 99 (1954).

\bibitem{chen-2020}
X. Chen and L. Jiang, Noise tolerance of Dicke states, Phys. Rev.
A {\bf 101}, 012308 (2020).

\bibitem{bausch-leditzky}
J. Bausch and F. Leditzky, Quantum codes from neural networks, New
J. Phys. {\bf 22}, 023005 (2020).

\bibitem{banik-2022}
G. L. Sidhardh, M. Alimuddin, M. Banik, Exploring super-additivity
of coherent information of noisy quantum channels through genetic
algorithms, arXiv:2201.03958.

\end{thebibliography}
\end{document}